\documentclass[%
 reprint,
 amsmath,amssymb,
 aps,showkeys,
]{revtex4-2}

\usepackage{graphicx}
\usepackage{dcolumn}
\usepackage{bm}

\begin{document}

\preprint{APS/123-QED}

\title{Molecular unfolding formulation with enhanced quantum annealing approach}

\author{Arit Kumar Bishwas}\email{arit.kumar.bishwas@pwc.com}
\affiliation{Innovation Hub, PricewaterhouseCoopers\\San Francisco, US}

\author{Arish Pitchai}\author{Anuraj Som}
\affiliation{Innovation Hub, PricewaterhouseCoopers\\Bangalore, India}

\date{\today}

\begin{abstract}
Molecular docking is a crucial phase in drug discovery, involving the precise determination of the optimal spatial arrangement between two molecules when they bind. In such analysis, the 3D structure of molecules is a fundamental consideration, involving the manipulation of molecular representations based on their degrees of freedom, including rigid roto-translation and fragment rotations along rotatable bonds, to determine the preferred spatial arrangement when molecules bind to each other. In this paper, quantum annealing based solution to solve Molecular unfolding (MU) problem, a specific phase within molecular docking, is explored and compared with a state-of-the-art classical algorithm named “GeoDock”. Molecular Unfolding focuses on expanding a molecule to an unfolded state to simplify manipulation within the target cavity and optimize its configuration, typically by maximizing molecular area or internal atom distances. Molecular Unfolding problem aims to find the torsional configuration that increases the inter-atomic distance within a molecule, which also increases the molecular area. Quantum annealing approach first encodes the problem into a Higher-order Unconstrained Binary Optimization (HUBO) equation which is pruned to an arbitrary percentage to improve the time efficiency and to be able to solve the equation using any quantum annealer. The resultant HUBO is then converted to a Quadratic Unconstrained Binary Optimization equation (QUBO), which is easily embedded on a D-wave annealing Quantum Processor.
\end{abstract}

\maketitle


\section{\label{sec1}Introduction}

Drug discovery is a fundamental aspect of modern medicine and healthcare, with wide-ranging impacts on human health, economies, and scientific progress. The process of discovering new drugs frequently extends over a decade and involves investments amounting to billions of dollars before a molecule can attain the status of a pharmaceutical drug \cite{cao2018potential}. A substantial share of these resources is allocated to the exploration of molecules demonstrating substantial medical efficacy against a particular disease \cite{sadybekov2023computational}. Conventionally, the initial step in the discovery process involves the creation of a repository of potential drug candidates, which is then subjected to screening for medicinal activity \cite{jayatunga2022ai}. The process of discovering new drugs encompasses a range of tasks conducted in vivo, in vitro, and in silico. Molecular docking, a task commonly carried out in silico, is one such component.

Molecular docking is a computational technique at the forefront of modern computer-assisted drug discovery, playing a pivotal role in understanding the interactions between small molecules and target proteins at the molecular level. The goal of molecular docking is to identify and optimize candidate compounds that can bind to the target protein with high affinity and specificity \cite{sabe2021current}. This powerful tool facilitates the prediction and analysis of binding affinities and poses between a small molecule (usually a potential drug candidate) and a biological target, typically a protein at the molecular level, guiding the design and optimization of novel drug candidates. By simulating the intricate rotations of atoms and molecules, molecular docking aids in the identification of potential drug leads, elucidating the structural basis of ligand-protein interactions, and accelerating the drug development process \cite{morris2008molecular}. Its versatility extends beyond pharmaceuticals to areas such as protein-protein interaction studies and the investigation of protein-ligand binding mechanisms in diverse biological contexts. The continual refinement and integration of various docking algorithms with experimental data continue to enhance our ability to discover and develop innovative therapeutics, making molecular docking an indispensable component of contemporary research in the life sciences. This article focuses on the Molecular unfolding (MU) problem, which is a stage in the process of molecular docking based on geometric approach.

Molecular Unfolding is a computationally intensive problem since it involves taking into consideration several atoms and degrees of freedom. Since docking time is an important factor to be considered and there are more than a million compounds in the search space, need for a faster computing technique has arisen. Dennard scaling and failure of Moore’s prediction paved way to the new computing paradigms \cite{bishwas2017quantum, pitchai2022quantum}. Quantum computing has rapidly progressed over the last few years and has the potential to impact several highly computationally intensive problems with the help of principles from quantum physics \cite{bishwas2020investigation, bishwas2020gaussian, heifetz2020quantum}. Quantum annealing computers are a specialized type of quantum computing technology designed to address optimization problems efficiently \cite{bishwas20204}. Unlike universal quantum computers that use quantum bits (qubits) for a wide range of computational tasks, quantum annealers utilize a unique approach inspired by quantum mechanics to solve specific types of problems, particularly optimization problems \cite{hauke2020perspectives}. Since MU problem involves maximization, the Molecular Unfolding problem could be potentially solved using quantum annealing. Specifically, it being an optimization problem, quantum annealing could help in exploring the vast energy landscape of solutions simultaneously of the different configurations of the molecule to maximize volume while being potentially faster. Thus, this article explores the applicability of quantum annealing in Molecular Unfolding. This capability can find valuable applications in drug discovery, promising innovative solutions to complex problems in these domains \cite{zinner2021quantum}.

The remainder of the paper is organized as follows. In section 2, the necessary preliminary topics such as molecular unfolding, quantum annealing and the mathematical formulation of the problem are introduced, while in section 3 the proposed methodology is discussed in detail. Section 4 includes the experimental findings related to the QPU (Quantum Processing Unit) embedding process, along with a comprehensive performance evaluation that compares the state-of-the-art classical method used. Finally, section 5 concludes the article followed by an Appendix.

\section{\label{sec2}Preliminaries}

\subsection{\label{subsec1}Molecular Unfolding}

Structure-based drug design leverages the three-dimensional geometric data to pinpoint appropriate ligands. Finding ligand poses for protein pockets is a critical aspect of structure-based drug design and molecular docking. This process involves the prediction of the three-dimensional orientation and binding conformation of a small molecule, known as a ligand, within the binding site or pocket of a target protein. The molecular docking algorithm is used to predict the optimal binding pose of each ligand within the protein's binding pocket. Ligand poses generated by the docking software are assessed and scored based on their predicted binding affinity and fitness within the protein pocket. 

Given that electrons within atoms exhibit mutual repulsion, their interaction plays a role in shaping molecules and influencing their reactivity. Consequently, by exerting control over the molecular configuration, it becomes feasible to make predictions regarding both the ligand's reactivity within a protein's pocket and the associated energy cost. This direct linkage between molecular conformation and binding affinity forms the basis for geometrical scoring functions. In our methodology, the docking process treats the pocket as an unchanging structure, while the ligand comprises a dynamic ensemble of atoms.

Typically, when employing a geometric approach in molecular docking, one can identify three primary stages: ligand expansion, initial positioning, and shape-refinement within the pocket \cite{gadioli2021tunable}. Molecular unfolding (MU) is a vital component of the ligand expansion phase, serving as a critical step in enhancing the precision of geometric docking procedures. The initial placement of the ligand, predefined beforehand, can potentially introduce a bias in terms of shape, which could impact the quality of the docking outcome. MU technique is being used to rectify this initial bias, effectively expanding the ligand into an unfolded conformation, thereby mitigating shape-related prejudices. 

\subsubsection{\label{subsubsec1}Problem Definition}

The goal of the Molecular Unfolding problem is to determine the unfolded conformation of the ligand, finding the torsional arrangement that maximizes the molecular volume. This volume is quantified by the total sum of internal distances between pairs of atoms within the ligand. Initially, the process begins with the folded molecule characterized by a specific set of atoms and its fixed and rotatable bonds. For each rotatable bond, a variable $t_i$ is assigned that represents the angle $\theta_i$ responsible for the rotation of the associated fragment.

The ordered set of torsional angles are represented as a single vector:

\begin{equation}
    \Theta = \left[ \theta_1, \theta_2, \dots, \theta_n \right].\label{eq1}
\end{equation}

where the rotational angle $\theta_i$, which can be assigned any value between $0$ and $2\pi$, belongs to the $i^{th}$ torsional bond $t_i$. Equation \eqref{eq1} also also signifies that the molecule is composed of $n$ rotational bonds. The primary objective of addressing the Molecular Unfolding (MU) problem is to determine the degree of rotation required for each torsional bond, leading to the transformation of the molecule into an unfolded state. The solution vector is represented as:

\begin{equation}
    \Theta^{unfold} = \left[ \theta_1^{unfold}, \theta_2^{unfold}, \dots, \theta_n^{unfold} \right].\label{eq2}
\end{equation}

Here, maximizing volume of a molecule involves expanding its spatial configuration by elongating the distances between its constituent atoms. This can be mathematically written as maximizing the following quantity:

\begin{equation}
    D\left(\Theta\right) = \sum_{\substack{u,v \in \mathcal{M} \\ u \neq v}} D_{uv}\left(\Theta\right)^2.\label{eq3}
\end{equation}

The distance between any two atoms $u$ and $v$ of the molecule $\mathcal{M}$ is represented as the function $D_{uv}\left(\Theta\right)$. Thus, $D\left(\Theta\right) $ is the sum of all square distances between all pairs of atoms present in $\mathcal{M}$. The reason for selecting this approach is that the objective function is straightforward, dependent solely on the molecular geometry.

\begin{figure*}
\includegraphics[width=0.6\textwidth]{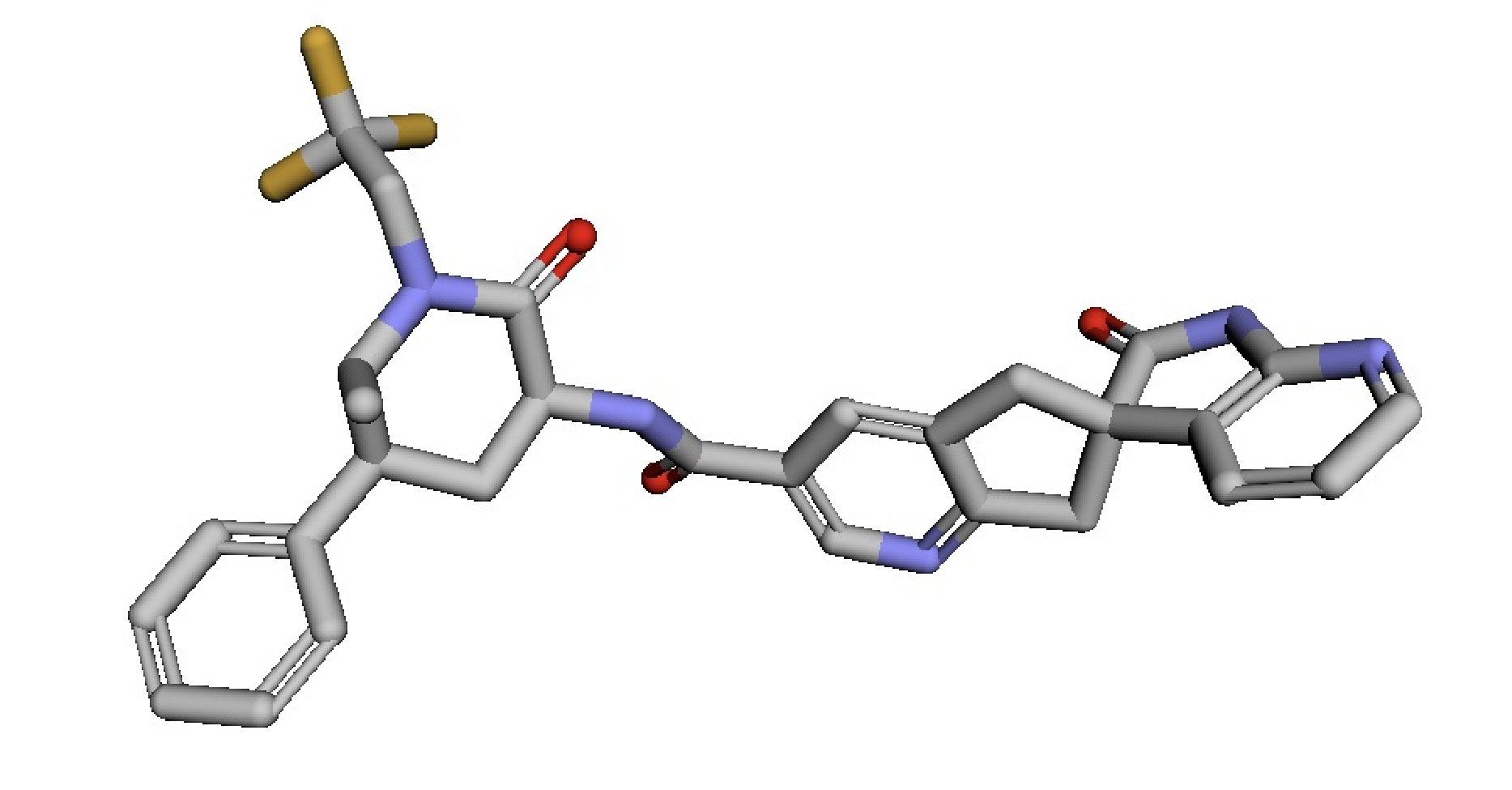}
\caption{\label{fig1} Representation of a ligand molecule.}
\end{figure*}

Since, the distance between the atoms that belong to the same rigid segment remains unchanged even after rotations, there's no need to compute every possible pair of distances (Refer Figure \ref{fig1}). Any pair of atoms is considered for distance calculation in equation \eqref{eq3} only if the shortest path between those atoms has at least one rotational bond. It can be easily observed that if the length of the shortest path between any pair of atoms is less than three, then the distance between them remains unchanged because of torsions. This leads to another important simplification of equation \eqref{eq3}. If an atom pair is included for the distance calculation, then a minimum of three edges is required in the shortest path between them.

Change in the coordinates of atoms due to rotations are calculated using rotation matrices $R\left(\theta_i\right)$, where $\theta_i$ is the angle in which $i^{th}$ torsional bond is being rotated. Mathematical construction of the $4\times4$ rotation matrices using the angle of rotation and the coordinates of extremes atoms are explained in detail in appendix \ref{secA1}. If a pair of atoms are separated by $k$ torsional bonds $\left(k > 1\right)$ in the order $t_{x_1}, t_{x_2}, \dots, t_{x_k}$, the rotation matrix $R\left(\Theta\right) = R\left(\theta_{x_1}, \theta_{x_2}, \dots, \theta_{x_k} \right)$ can be derived from the following equation:

\begin{equation}
    R\left(\Theta\right) = R\left(\theta_{x_1}, \theta_{x_2}, \dots, \theta_{x_k} \right) = R\left(\theta_{x_1}\right) \times R\left(\theta_{x_2}\right) \times \dots \times R\left(\theta_{x_k}\right).\label{eq4}
\end{equation}

When one or more bonds between a pair of atoms are rotated, one of the atoms is fixed and the other transforms relative to the fixed atom. Let's assume a set bond between atoms $u$ and $v$ is rotating, the initial coordinates of one $\vec{u}_0$ is fixed and the initial coordinates of the other $\vec{v}_0$ gets modified as follows:

\begin{align}
    \vec{u}_{new} = \vec{u}_0 \nonumber \\
    \vec{v}_{new} = R\left(\Theta\right)\vec{v}_0. \label{eq5}
\end{align}

Based on equation \eqref{eq5} and the fact that Euclidean distance is the metric used in the distance function, the single distance function $D_{uv}\left(\Theta\right)^2$ is represented as:

\begin{equation}
    D_{uv}\left(\Theta\right)^2 = \left\| \vec{u}_0 - R\left(\Theta\right)\vec{v}_0 \right\|^2.\label{eq6}
\end{equation}

The following section is dedicated to introduce the quantum approach, that is being used widely to solve business applications, called the quantum annealing technique.

\subsection{\label{subsec2}Quantum Annealing}

Quantum annealing is a quantum computing technique that utilizes the principles of quantum superposition, tunneling, and thermal fluctuations to explore and potentially find optimal solutions to complex optimization and sampling problems. It harnesses the unique properties of quantum bits or qubits to explore vast solution spaces efficiently, potentially offering significant advantages over classical computing methods. 

One of the benefits of quantum annealing is the ability of qubits to exist in superposition, thus making them capable of representing multiple possibilities simultaneously. Additionally, quantum annealing exploits quantum tunneling, a phenomenon that enables quantum systems to overcome energy barriers and explore different regions of the energy landscape. Consequently, thermal fluctuations are introduced to mimic the annealing process, causing the system to gradually decrease in temperature. This temperature reduction encourages the quantum system to settle into a low-energy state, ideally locating the lowest energy state that corresponds to the optimal solution of the optimization problem. 

In Quantum Annealing, the solution to a problem is arrived at by going through several steps. It starts with the initial Hamiltonian, which is used to portray the landscape of the problem and moves increasingly towards the final Hamiltonian according to predefined annealing functions while minimizing the energy of the system. This can be done both with the help of a pure quantum or even a hybrid approach as well. 

Quantum Annealing is a form of Adiabatic Quantum Computing, where the conditions of adiabacity are not met, resulting in a heuristic variational quantum algorithm. Thus, it can be used to find the ground state of Ising models, a known NP-hard task. Using Ising and Quadratic Unconstrained Binary Optimization (QUBO) forms, several optimization and sampling problems can be brought into forms suitable for solving through quantum annealing. 

\subsection{\label{subsec3}Higher order Unconstrained Binary Optimization (HUBO)}

\begin{figure*}
\includegraphics[width=0.6\textwidth]{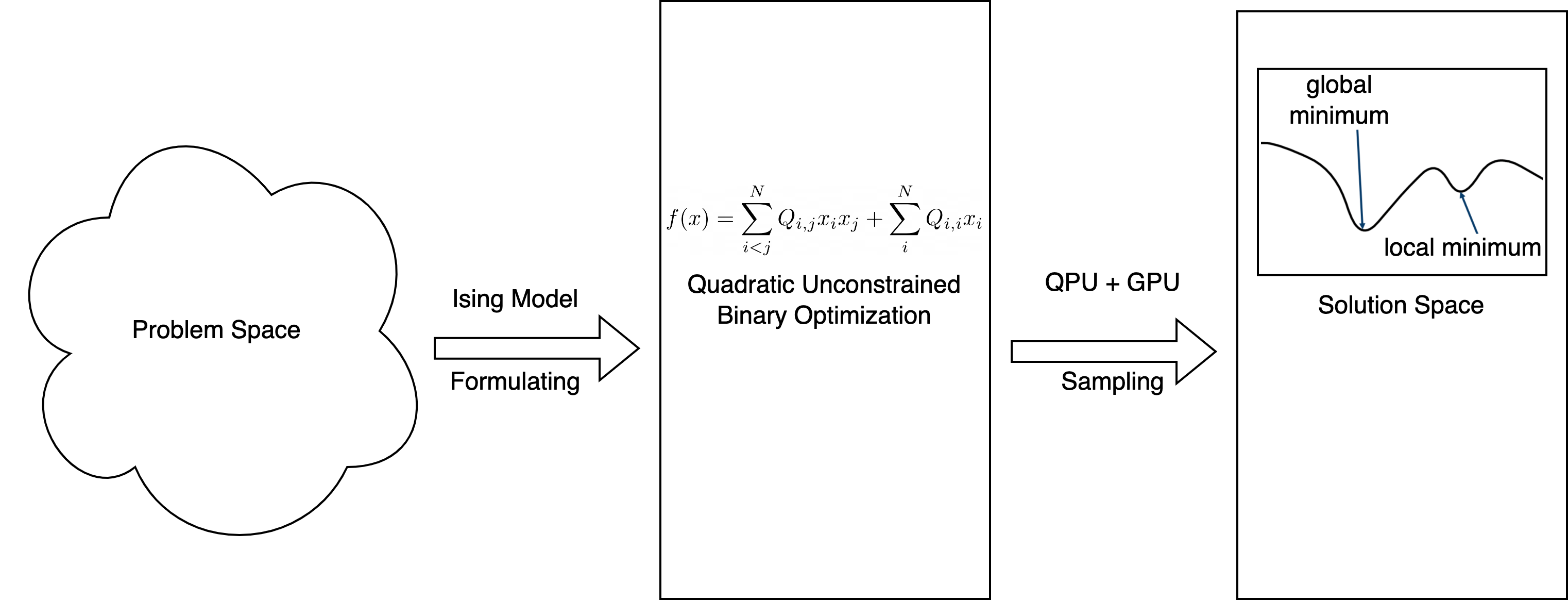}
\caption{\label{fig2} Solving an optimization problem using a quantum annealer.}
\end{figure*}

Major step in solving any optimization problem using quantum computer is to formulate the problem as an objective function. Objective functions (or cost functions) are mathematical representations of the optimization problem at hand. In the realm of quantum computing, these are typically expressed as quadratic models (QUBOs), where optimal solutions to the problems they depict correspond to the lowest energy. The problem in QUBO form is then submitted to a quantum sampler to find the solution as illustrated in Figure \ref{fig2}.

A generalized form of a QUBO formulation is defined as:

\begin{equation}
    Q(x) = \sum_i a_i x_i + \sum_{i>j} b_{i,j} x_i x_j.\label{eq7}
\end{equation}

where $x_i \in \{0, 1\}$ are the binary variables used to encode the solution while $a_i$ and $b_{i,j}$ are the coefficients which are being constructed based on the problem description \cite{kochenberger2014unconstrained}. This problem is formulated in a manner where the quantum solution $x$, represented as a binary vector, seeks to minimize $Q\left(x\right)$.

Not every problem can be expressed as a QUBO equation. Numerous real-world problem instances exhibit a higher-order nature by default. For complex problems such as Molecular Unfolding, which involve interactions among more than two variables, the process of conversion to QUBO becomes challenging. Such problems may be converted to a Higher-order Unconstrained Binary Optimization (HUBO) equation as represented below:

\begin{equation}
    Q_{HUBO}(x) = \sum_i a_i x_i + \sum_{i,j} b_{i,j} x_i x_j + \sum_{i,j,k} c_{i,j,k} x_i x_j x_k + \dots.\label{eq8}
\end{equation}

Generalizeation of QUBO is the HUBO representation, which accommodates not only quadratic terms but also higher-order terms within the objective function \cite{domino2022quadratic}. Compressed quadratization of higher order binary optimization equations makes it easy to solve complex problems using quantum annealers\cite{mandal2020compressed}. 

Formulating the problem of maximizing the molecule volume into a HUBO starts by defining the binary variables. Solution space of this problem consists of different angle of rotations for torsional bonds. The initial step in HUBO formulation involves discretizing any torsional angle $\theta_i$ into $d$ distinct values with equally spaced intervals as shown below:

\begin{equation}
    \theta_i = \left[ \theta_i^1, \theta_i^2, \dots, \theta_i^d \right].\label{eq9}
\end{equation}

For example, if $d =  4$, then $\theta_i \in \{ 0, \frac{\pi}{2}, \pi, \frac{3\pi}{2} \}$. Discretization of the angles leads to the discretization of their respective $\sin$ and $\cos$ values as represented below:

\begin{equation}
    \sin{\theta_i} = \left[ \sin{\theta_i^1}, \sin{\theta_i^2}, \dots, \sin{\theta_i^d} \right].\label{eq10}
\end{equation}

\begin{equation}
    \cos{\theta_i} = \left[ \cos{\theta_i^1}, \cos{\theta_i^2}, \dots, \cos{\theta_i^d} \right].\label{eq11}
\end{equation}

The formulated mathematical model is expected to assign only one value out of $d$ for a torsional angle $\theta_i$. This constraint is named as $\text{hard constraint}$. To achieve this, a set of $d$ binary variables $x_{ik}$, with $1 \leq k \leq d$ and $x_{ik} \in \{0, 1\}$, are associated with each torsional bond $t_i$. One is assigned to $x_{ik}$, if the $k^{th}$ value is is assigned to $\theta_i$ and rest of them are assigned with zero. Hard constraint can also be expressed as:

\begin{equation}
    x_{ik} = 
\begin{cases}
    1,  & \text{if } \theta_i = \theta_i^k;\\
    0,  & \text{otherwise}.
\end{cases}.\label{eq12}
\end{equation}

This one-hot encoding approach signifies that only a single binary variable can be assigned a value of one and as a result of which their sum will also be equal to one.

\begin{equation}
    \sum_{k=1}^d x_{ik} = 1.\label{eq13}
\end{equation}

Currently the quantum annealers are capable of only minimizing the objective function submitted. To solve this one-hot constraint by minimization, equation \eqref{eq13} is re-written as an equivalent objective function:

\begin{equation}
    \min \left( \sum_{k=1}^d x_{ik} - 1 \right)^2.\label{eq14}
\end{equation}

If there are $n$ rotational bonds present in the ligand molecule, then the mathematical modeling of the hard constraint is obtained from equation \eqref{eq14} and it is represented below as:

\begin{equation}
    \min \sum_{i=1}^n \left( \sum_{k=1}^d x_{ik} - 1 \right)^2.\label{eq15}
\end{equation}

The primary objective of solving the $\text{MU}$ problem is to maximize the Euclidean distance between all pairs of atoms as defined in equation \eqref{eq3}. This equation can also be called as the $\text{distance constraint}$. The function $D_{uv}\left(\Theta\right)^2$ of equation \eqref{eq3} is constructed based on the rotated coordinates of atoms $u$ and $v$, which, in themselves, are the outcome of rotational transformations as mentioned in equation \eqref{eq5}. It is known that the problem is to select one of the angles from a set of $d$ values for any $\theta_i$. To achieve this, $\sin{\theta_i}$ in rotation matrix $R\left( \theta_i \right)$ is replaced with the sum of $d$ different $\sin$ values corresponding to $d$ different angles, multiplied with their respective binary variables $x_{ik}$, which can be expressed as:

\begin{equation}
    \sum_{k=1}^d x_{ik} * \sin{\theta_i^k}.\label{eq16}
\end{equation}

Similarly, $\cos{\theta_i}$ is replaced with the following expression:

\begin{equation}
    \sum_{k=1}^d x_{ik} * \cos{\theta_i^k}.\label{eq17}
\end{equation}

Hence, the rotation matrix $R\left( \theta_i \right)$ is dependant on all the necessary binary variables $x_{ik}$, used to depict the angle $\theta_i$ as described below:

\begin{equation}
    R\left( \theta_i \right) = R\left( x_{i1}, x_{i2}, \dots, x_{id} \right).\label{eq18}
\end{equation}

As mentioned earlier in this section, quantum solvers are capable of only minimizing the objective function. So, it is necessary to multiply the distance constraint by $-1$ in order to convert the maximization problem to a minimization task. Thus, the final $\text{HUBO}$ equation of the $\text{MU}$ problem, which has both hard and distance optimization constraint, is formulated below as:

\begin{equation}
    O_{MU}\left(x_{ik}\right) = A_{const} * \sum_{i=1}^n \left( \sum_{k=1}^d x_{ik} - 1 \right)^2 - \sum_{\substack{u,v \in \mathcal{M} \\ u \neq v}} D_{uv}\left(\Theta\right)^2.\label{eq19}
\end{equation}

Here, value of $A_{const}$ decides the strength of the hard constraint. Usually $A_{const}$ is selected as a value greater than the largest coefficient in the HUBO. The following section explains the procedure of implementin the solution in detail.

\section{Proposed Methodology}\label{sec3}

\begin{figure*}
\includegraphics[width=0.6\textwidth]{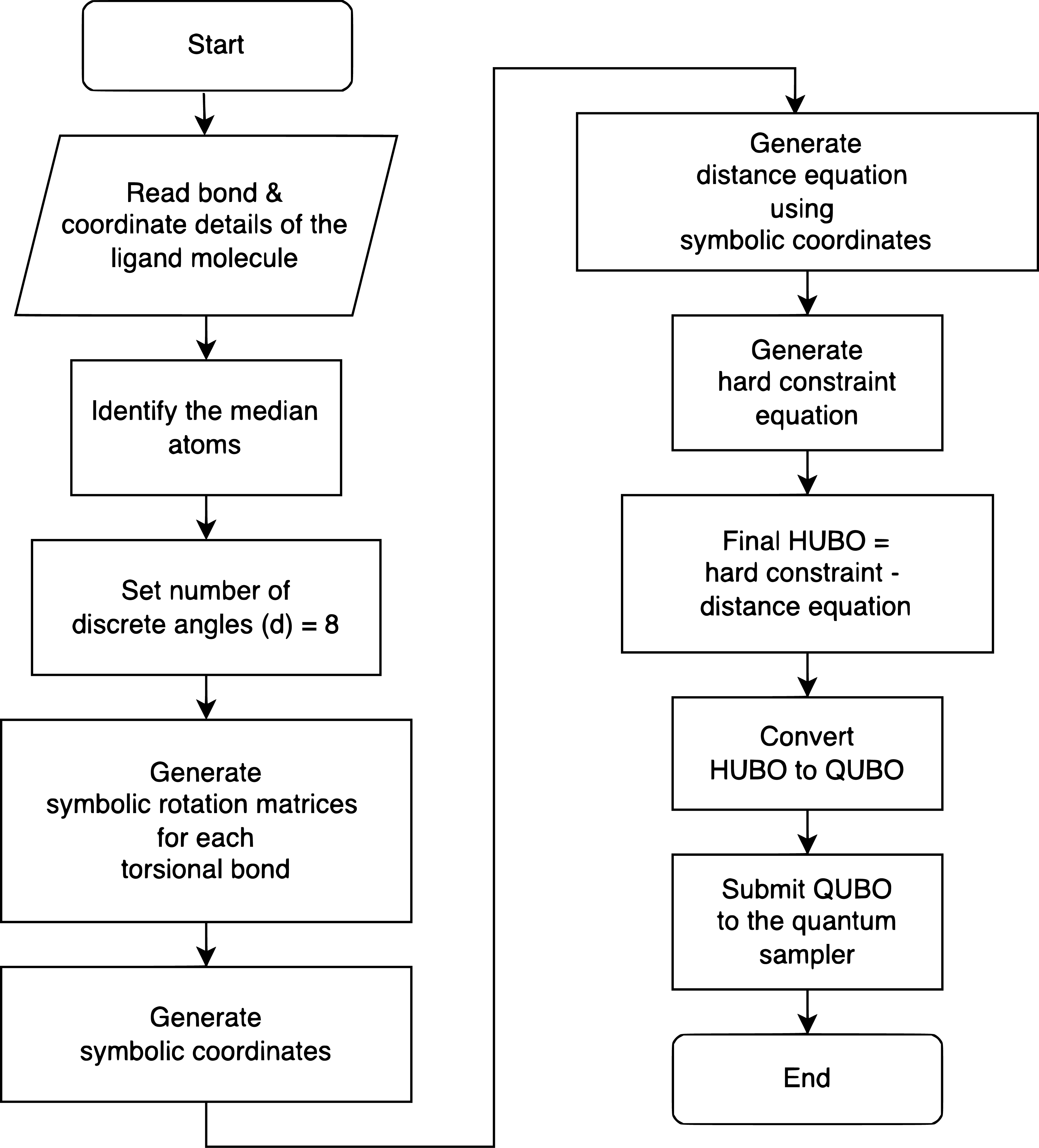}
\caption{\label{fig3} Framework of quantum molecular unfolding.}
\end{figure*}

The ligand dataset being used for the experiments was obtained from Protein Data Bank (PDBbind) \cite{wwpdb2019protein} and comprises molecules with atom counts ranging from 20 to over 120, as well as torsional counts of up to around 50. It is evident that a majority of the ligands in the database possess between 5 and 10 torsional bonds, with a peak in the number of atoms occurring in the range of approximately 40 to 50. In this article, experiments are focused on the molecules with up to 13 rotatable bonds and up to 50 atoms. Each molecule undergoes a series of steps before estimating the torsional configuration as shown in Fig. \ref{fig3}. Different phases of the algorithm are described below.

\subsection{Preprocessing phase}

During the initial preprocessing step, we parse the molecule file in 'mol' or 'mol2' format and generate a three-dimensional structural representation. In this 3D visualization, we distinguish single bonds as torsional bonds. Terminal hydrogen bonds are excluded from consideration, as they have no relevance to problem resolution. To streamline problem complexity, we limit our analysis to a maximum of two atoms within any rigid fragment. Given the computational expense associated with formulating the HUBO problem, this significant simplification accelerates the classical phase of our solution. 

\subsection{Hard constraint preparation}

Following the initial molecule pre-processing, the overall count of torsional bonds is determined. From the literature \cite{mato2022quantum}, it is observed that eight discrete angles are appropriate to consider for each rotation. Consequently, the total number of variables constituting the hard constraint equation is calculated as the product of the total rotatable bonds and the number of discrete angles. 

\subsection{Symbolic transformation of coordinates}

As mentioned earlier, only a subset of atoms, named as median atoms, are being used in the HUBO formulation. In order to construct the distance constraint $D\left(\Theta\right)$ using the median atoms, each coordinate is converted as a function of the one-hot variable $x_{ik}$ as explained in equation \eqref{eq12}. Since, the number of terms in each symbolic representation of the median atom coordinates are exponentially growing, a thresholding mechanism is adopted to eliminate insignificant (too small and too large) coefficients \cite{sax2020approximate}.

\subsection{Extracting solution from final HUBO}

The final HUBO, as represented in equation \eqref{eq19}, is the sum of hard constraint multiplied with a constant and the distance constraint multiplied with one. Since, the number of terms keeps increasing in this concatenation phase, there is an increase in memory to compute the equation. So, thresholding of insignificant terms is carried out again as suggested in \cite{sax2020approximate}. Implementing this threshold approximation had the result of speeding up the HUBO construction and reducing the number of terms within the Quadratic Unconstrained Binary Optimization (QUBO) equation to align with the specific constraints and limitations, like the available number of qubits and the connectivity of the problem, of the quantum annealing (QA) hardware. The mathematically approximated QUBO is then embedded into a quantum annealing processor to obtain multiple solutions. Few, around ten, of the low energy feasible solutions are evaluated and the solution with the highest volume gain is considered as the quantum solution to the problem. 

\section{Results and Discussion}\label{sec4}

Nine distinct molecules, rotations ranging from five to thirteen, are selected from the protein data bank (PDB) database. The classical computations that formulates QUBO for the given molecules are performed in a standard NC24ads A100 v4 virtual machine which is comprised of 24 vCPUs, 220 GiB memory and a GPU with 80 GiB memory. The quantum annealing experiments are minor-embedded on the structured quantum samplers with D-wave's Pegasus topology graph, which are oriented vertically and horizontally. Geodock is the state-of-art classical algorithm used to benchmark the results. The Geodock search employs a greedy approach, involving the gradual rotation of an increasing number of bonds. Various measures for the quality of the solution are considered such as execution time, gain in the volume and time to solution.

\subsection{Volume gain in time with different torsionals}
Finding the percentage of increase in the volume as a result of changes conformation in a limited time duration is the primary objective of the problem. The results in Table \ref{tab1} and Table \ref{tab2} compares the execution times and volume gain percentages respectively between the quantum annealers and the classical GeoDock algorithm. 

\begin{table}[b]
\caption{\label{tab1} Number of rotations Vs Execution Time.}
\begin{ruledtabular}
\begin{tabular}{ccc}
No.       & Quantum   & Classical \\
of        & Execution & Execution \\
rotations & Time (s)  & Time (s)  \\
\hline
5   & 63.44     & 165.04 \\
6   & 72.9      & 199.988 \\
7   & 280.16    & 358.55 \\
8   & 338.82    & 1253.34 \\
9   & 926.59    & 2568.98 \\
10  & 2967.76   & 8958.76 \\
11  & 34814.39  & 57595.22 \\
12  & 29128.61  & 79192.113 \\
13  & 70343.81	& 177716.882 \\
\end{tabular}
\end{ruledtabular}
\end{table}

It is evident from the empirical results that the quantum solutions are faster than the classical method in all the cases. The lowest and the highest speed ups are found to be 39.55\% for a molecule with 11 rotations and 72.97\% for an 8 rotations molecule. An average speedup of 57.1\% is the strongest evidence of quantum supremacy.

\begin{table}[b]
\caption{\label{tab2} Number of rotations Vs Volume Gain percent.}
\begin{ruledtabular}
\begin{tabular}{ccc}
No.       & Quantum     & Classical   \\
of        & Volume Gain & Volume Gain \\
rotations & Percent     & Percent     \\
\hline
5   &4.99       &5.945 \\
6	&10.667   &11.421 \\
7	&30.769   &34.526 \\
8	&20.001   &35.272 \\
9	&49.388   &62.076 \\
10	&59.894  &64.113 \\
11	&44.11   &68.379 \\
12	&43.388  &64.005 \\
13	&202.242 &271.132 \\
\end{tabular}
\end{ruledtabular}
\end{table}

\begin{figure*}
\includegraphics[width=0.6\textwidth]{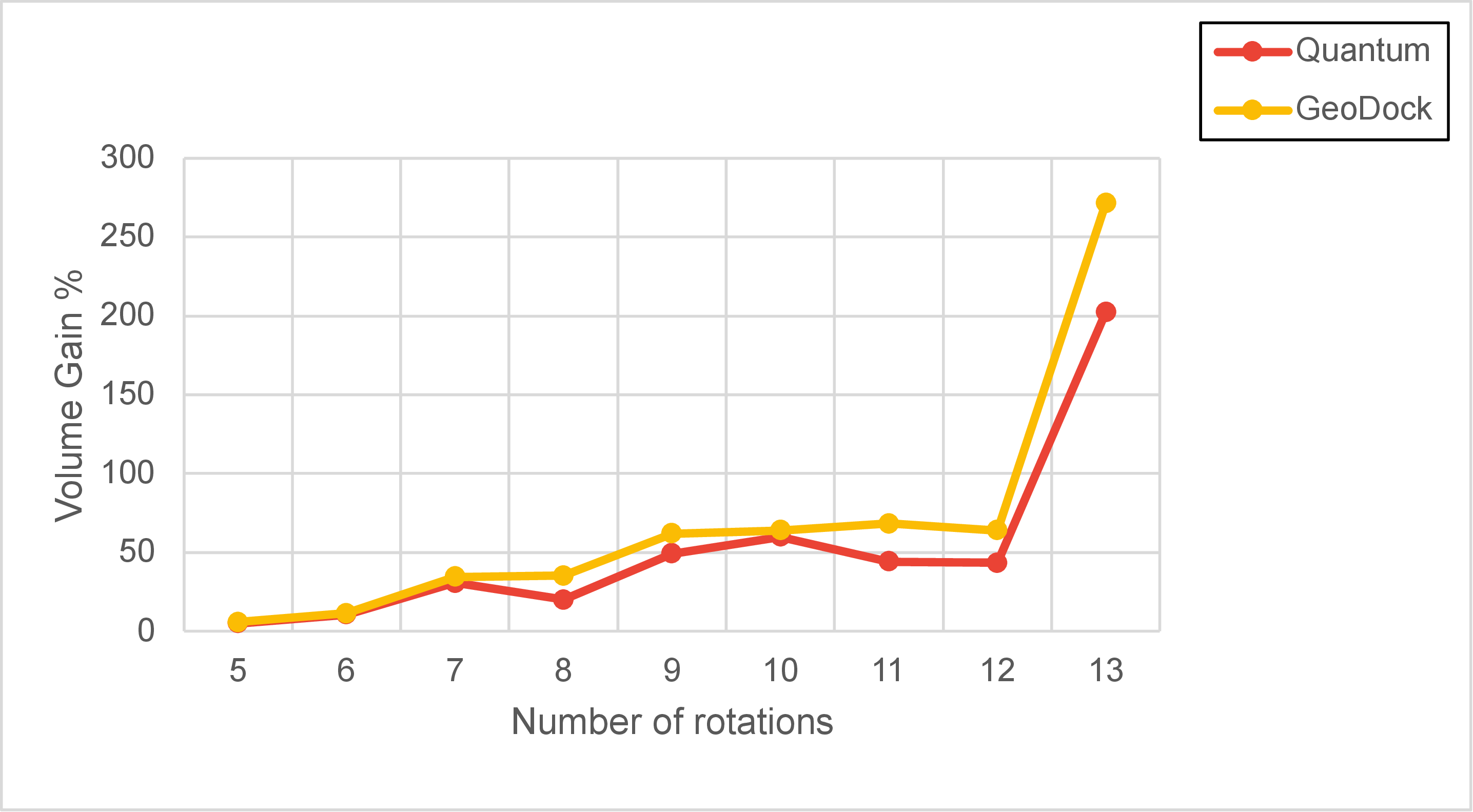}
\caption{\label{VGvsRots} Variation in Volume Gain(VG\%) with change in the Number of rotations.}
\end{figure*}

Due to the loss of large number of significant terms in the HUBO during thresholding processes, achieving better volume gain percentage is difficult for the quantum procedure as shown in Fig. \ref{VGvsRots}. The lowest percentage of advantage in volume gain being 6.58\% for a 10 rotational molecule and highest being 35.49\% for a molecule with 11 torsional angles. Since, the average percentage of advantage due to the application of classical GeoDock over the quantum technique is only 21.89\%, the proposed quantum algorithm can be claimed to give near optimal solutions similar to the greedy classical method.

\subsection{Influence of threshold value in volume gain}

\begin{figure*}
\includegraphics[width=0.6\textwidth]{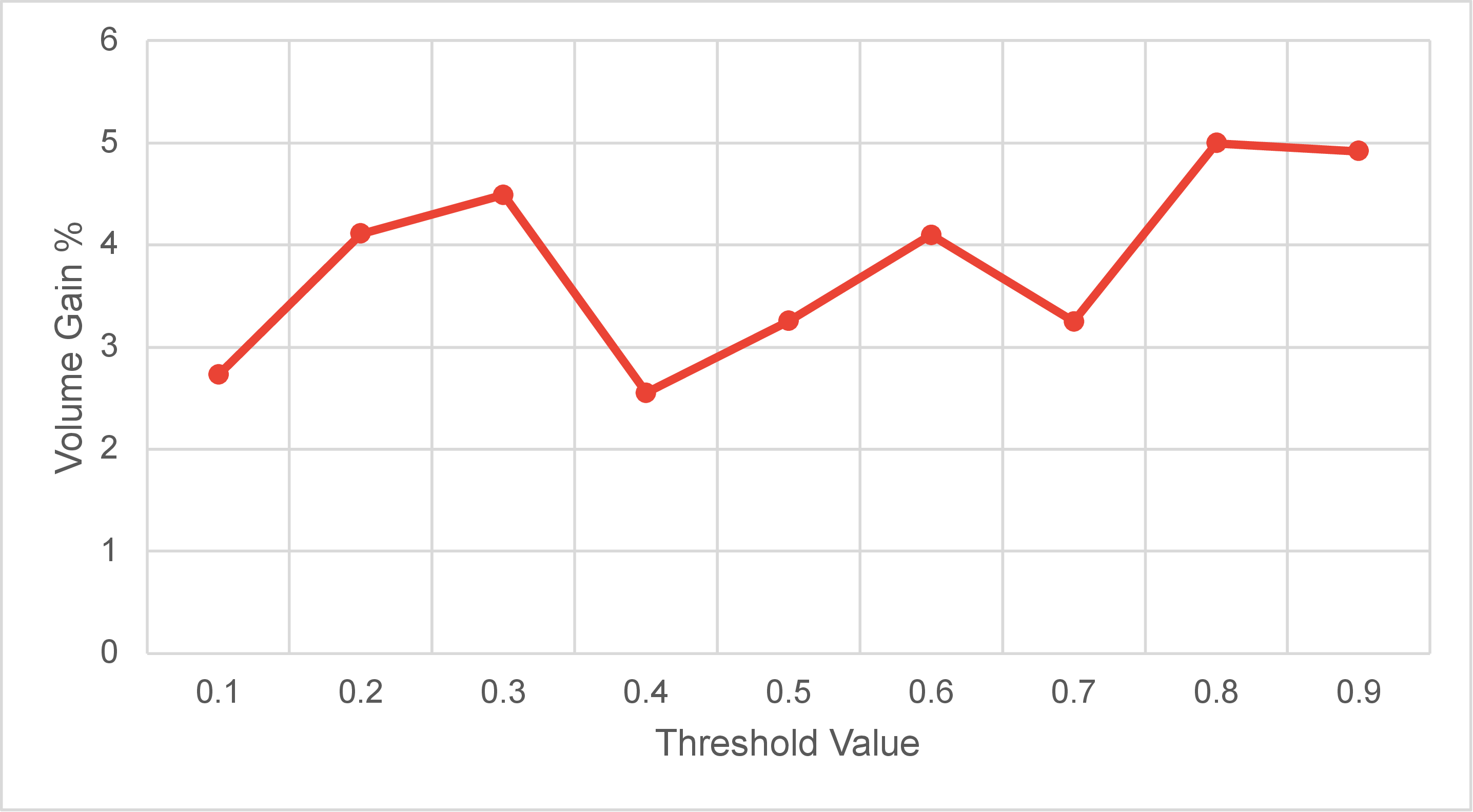}
\caption{\label{Lumateperone_VGvsTV} Variation in Volume Gain(VG\%) with change in the threshold value for Lumateperone.}
\end{figure*}

\begin{figure*}
\includegraphics[width=0.6\textwidth]{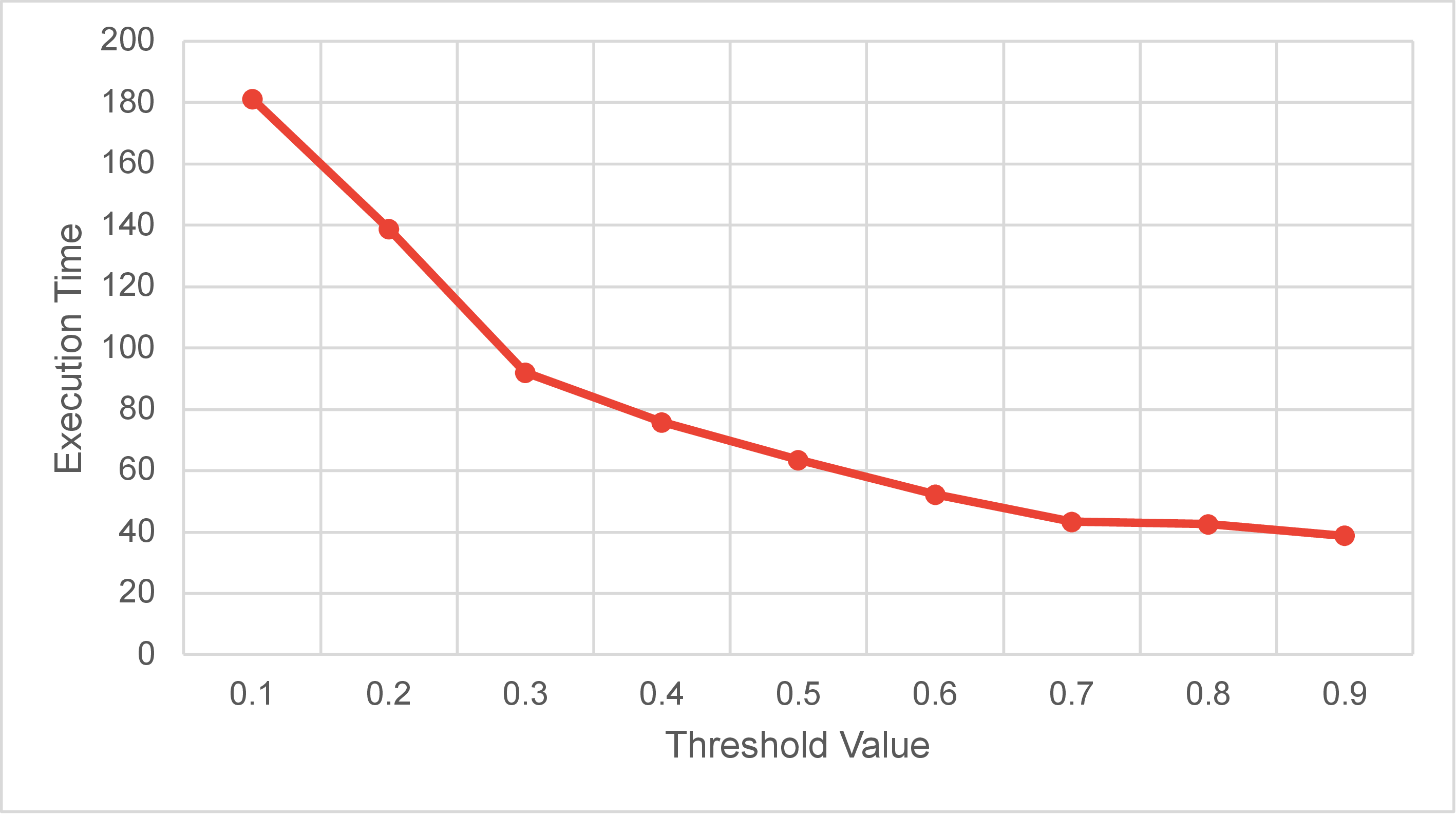}
\caption{\label{Lumateperone_TTSvsTV} Variation in Time-to-Solution(TTS) with change in the Threshold Value for Lumateperone.} 
\end{figure*}

The molecule lumateperone which has five single rotational bond was investigated with different threshold values ranging from 0.1 till 0.9 and the results are provided in Table \ref{tab3} and Fig. \ref{Lumateperone_VGvsTV} displays the corresponding trend in a line graph. The pearson correlation between the threshold values and the volume gain percent is found to be 0.53, which indicates a moderate positive correlation. It can also be seen from Fig. \ref{Lumateperone_TTSvsTV} that as the threshold value increases, the execution time on the classical machine is decreasing because of the reduction in the number of terms to formulate the final HUBO. So, it always recommended to select a threshold value more than 0.5 to achieve efficiency in time and better quality in solution.

\begin{table}[b]
\caption{\label{tab3} Volume Gain Vs Time Taken Vs Threshold Change.}
\begin{ruledtabular}
\begin{tabular}{ccc}
Threshold  & Quantum     & Execution \\
Value      & Volume Gain & Time      \\
Applied    & Percent     & (s)       \\
\hline
0.1 &2.728  &181.25 \\
0.2 &4.108  &138.84 \\
0.3 &4.491  &92.06 \\
0.4 &2.551  &75.81 \\
0.5 &3.254  &63.44 \\
0.6 &4.096  &52.18 \\
0.7 &3.251  &43.3 \\
0.8 &4.998  &42.56 \\
0.9 &4.917  &38.73 \\
\end{tabular}
\end{ruledtabular}
\end{table}

\subsection{Time taken for same volume gain percentage}
\begin{figure*}
\includegraphics[width=0.6\textwidth]{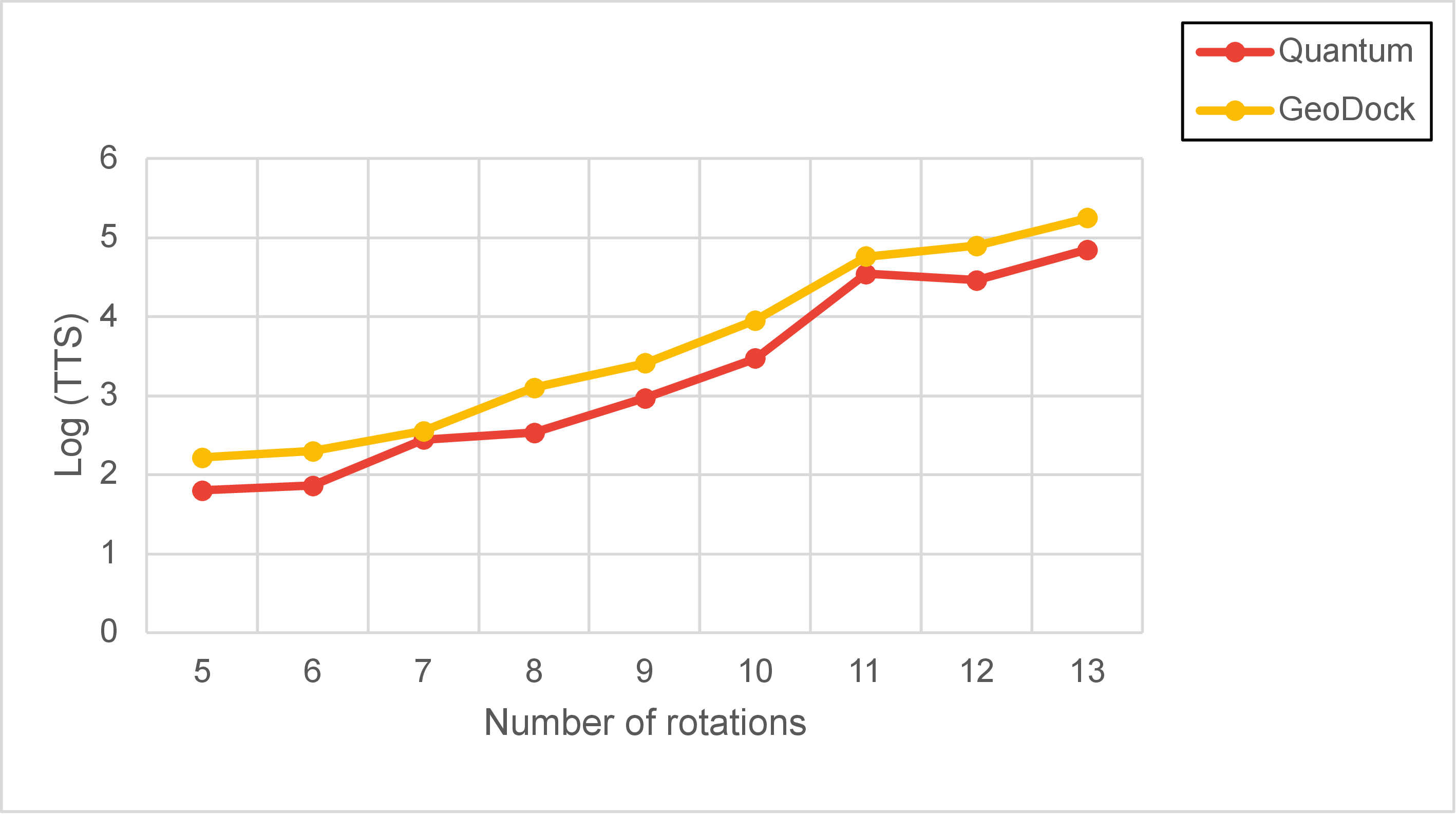}
\caption{\label{TTSvsRots} Variation in Time-to-Solution(TTS) in seconds with change in the Number of rotations.} 
\end{figure*}

\begin{figure*}
\includegraphics[width=0.6\textwidth]{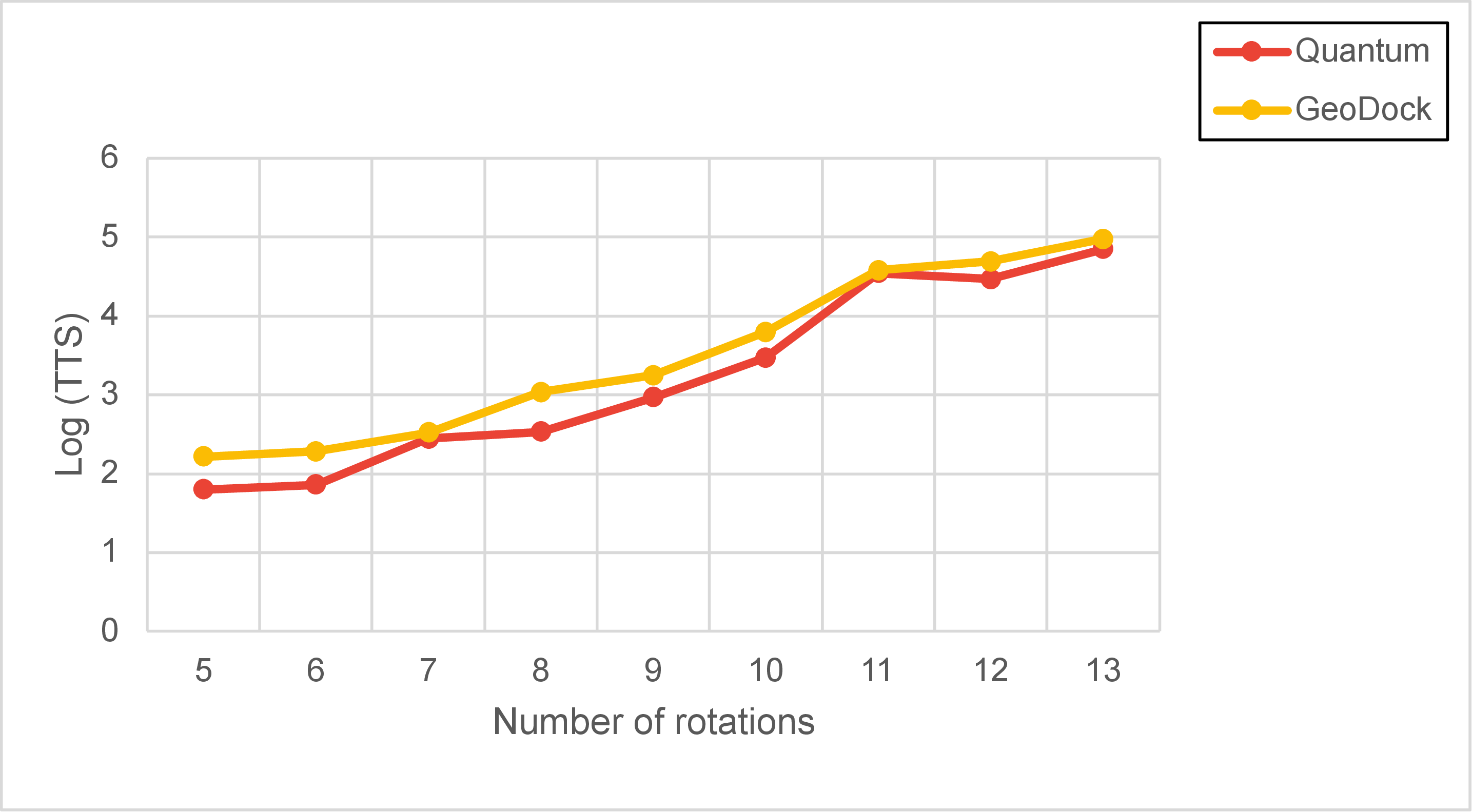}
\caption{\label{VGsame_TTSvsRots} Variation in Time-to-Solution(TTS) in seconds with change in the Number of rotations for similar Volume Gain \%.} 
\end{figure*}

Even-though GeoDock proves to give better volume gain in all the instances tested (Refer Fig. \ref{VGvsRots}), the ratio between the volume gain percentage and the time taken, as represented in Fig. \ref{TTSvsRots}, shows that quantum works faster than its classical counterpart. An experiment to find the time taken for GeoDock algorithm to achieve the volume gain percent reached by the QPUs is conducted and the results are given in Table \ref{tab4}. It can also be seen from Fig. \ref{VGsame_TTSvsRots} that when the time taken by classical computer to achieve the volume gain percentage is compared with the quantum algorithm, the classical algorithm fails to outperform. The percentage of quantum speed up achieved is varies from 8.6\% (for 11 rotation molecule) to 62.2\% (for 6 rotations) with a mean speed up of 42.5\%.

\begin{table}[b]
\caption{\label{tab4} Quantum Vs Geo dock execution time for same volume gain percentage.}
\begin{ruledtabular}
\begin{tabular}{ccc}
No.       & Quantum   & Classical \\
of        & Execution & Execution \\
rotations & Time (s)  & Time (s)  \\
\hline
5	&63.44	&164.1956476 \\
6	&72.9	&192.6541317 \\
7	&280.16	&331.6056295 \\
8	&338.82	&1076.262643 \\
9	&926.59	&1760.676695 \\
10	&2967.76	&6231.211948 \\
11	&34814.39	&38084.9917 \\
12	&29128.61	&49278.74632 \\
13	&70343.81	&94151.09457 \\
\end{tabular}
\end{ruledtabular}
\end{table}

\section{Conclusion}\label{sec5}
In this article, a detailed examination of the practical execution of the Molecular Unfolding problem has been provided. Implementation of the a quantum annealing approach to unfold the ligand molecule which helps to improve the efficiency of geometric molecular docking process is evaluated, aiming to explore potential industry-relevant applications on currently accessible quantum devices.
The benchmarks were conducted on nine distinct categories of problem instances to emphasize the varied challenges inherent in achieving a reliable and robust evaluation. Additionally, comparison between the proposed quantum approach and Geo Dock, the state-of-art classical method, is presented. The proposed solution begins by identifying the rotatable bonds, which serve as the parameters for the MU problem. After rewriting their discrete rotations using one-hot encoding through the introduction of binary variables, the overall sum of internal atomic distances can be formulated as a HUBO. Given the substantial number of terms in the polynomial generated, various simplifications such as picking median atoms, pruning the coefficients of symbolic coordinates and pruning the final HUBO terms are implemented. Since, pruning most of the terms in the formulated equation tends to reduce the quality of optimization, the future research direction in solving this MU problem could focus on using more qubits and less pruning of the monomials.

\textbf{Disclaimer:}
As of the time of this research, PwC does not have any joint business relationship (JBR) with D-Wave.

\appendix*

\section{Rotation Matrix}\label{secA1}

This appendix section presents the rotation matrix that is utilized to adjust the position of each atom in a rotatable fragment when the coordinates of the corresponding rotatable bond is known. This matrix serves as the foundation for the formulation discussed in this article. It is generated by taking into account the rotation angle $\theta$ and the coordinates of the two atoms $a_1 = \left( x_1, y_1, z_1 \right)$ and $a_2 = \left( x_2, y_2, z_2 \right)$ connected by the torsional bond. The matrix is constructed based on the rotation occurring around an arbitrary line that passes through the rotatable bond and has a direction vector from $a_1$ to $a_2$ \cite{van2015essential}.


Let us define $l_x = x_2 - x_1$, $l_y = y_2 - y_1$, and $l_z = z_2 - z_1$, as the $x$,$y$ and $z$ components of the direction vector from $a_1$ to $a_2$ respectively, while $l = \sqrt{(l_x^2 + l_y^2 + l_z^2)}$ as the length of the vector. In order to display the matrix in a limited space, $\sin{\theta}$ is shortened as $s_{\theta}$ and $\cos{\theta}$ is represented as $c_{\theta}$. We can now write the rotation matrix $R(\theta)$ as follows:


\begin{widetext}
\begin{equation}
\left[\begin{array}{cccc}
\frac{(l_x^2 + (l_y^2 + l_z^2)c_{\theta})}{l^2} & \frac{(l_x l_y(1-c_{\theta}) - l_zls_{\theta})}{l^2} & \frac{(l_x l_z(1-c_{\theta}) + l_yls_{\theta})}{l^2} & 
\frac{((x_1(l_y^2 + l_z^2)-l_x(y_1l_y + z_1l_z))(1-c_{\theta})+(y_1l_z - z_1l_y)ls_{\theta})}{l^2}	\\

\frac{(l_x l_y(1-c_{\theta}) + l_zls_{\theta})}{l^2} & \frac{(l_y^2 + (l_x^2 + l_z^2)c_{\theta})}{l^2} & \frac{(l_y l_z(1-c_{\theta}) - l_xls_{\theta})}{l^2} & 
\frac{((y_1(l_x^2 + l_z^2)-l_y(x_1l_y + z_1l_z))(1-c_{\theta})+(z_1l_x - x_1l_z)ls_{\theta})}{l^2}\\

\frac{(l_x l_z(1-c_{\theta}) - l_yls_{\theta})}{l^2} & \frac{(l_y l_z(1-c_{\theta}) + l_xls_{\theta})}{l^2} & \frac{(l_z^2 + (l_x^2 + l_y^2)c_{\theta})}{l^2} & 
\frac{((z_1(l_x^2 + l_y^2)-l_z(x_1l_x + y_1l_y))(1-c_{\theta})+(x_1l_y - y_1l_x)ls_{\theta})}{l^2}\\

0 & 0 & 0 & 1
\end{array}\right]
\end{equation}
\end{widetext}

\nocite{*}

\bibliography{apssamp}

\providecommand{\noopsort}[1]{}\providecommand{\singleletter}[1]{#1}%
\begin{thebibliography}{21}%
\makeatletter
\providecommand \@ifxundefined [1]{%
 \@ifx{#1\undefined}
}%
\providecommand \@ifnum [1]{%
 \ifnum #1\expandafter \@firstoftwo
 \else \expandafter \@secondoftwo
 \fi
}%
\providecommand \@ifx [1]{%
 \ifx #1\expandafter \@firstoftwo
 \else \expandafter \@secondoftwo
 \fi
}%
\providecommand \natexlab [1]{#1}%
\providecommand \enquote  [1]{``#1''}%
\providecommand \bibnamefont  [1]{#1}%
\providecommand \bibfnamefont [1]{#1}%
\providecommand \citenamefont [1]{#1}%
\providecommand \href@noop [0]{\@secondoftwo}%
\providecommand \href [0]{\begingroup \@sanitize@url \@href}%
\providecommand \@href[1]{\@@startlink{#1}\@@href}%
\providecommand \@@href[1]{\endgroup#1\@@endlink}%
\providecommand \@sanitize@url [0]{\catcode `\\12\catcode `\$12\catcode
  `\&12\catcode `\#12\catcode `\^12\catcode `\_12\catcode `\%12\relax}%
\providecommand \@@startlink[1]{}%
\providecommand \@@endlink[0]{}%
\providecommand \url  [0]{\begingroup\@sanitize@url \@url }%
\providecommand \@url [1]{\endgroup\@href {#1}{\urlprefix }}%
\providecommand \urlprefix  [0]{URL }%
\providecommand \Eprint [0]{\href }%
\providecommand \doibase [0]{https://doi.org/}%
\providecommand \selectlanguage [0]{\@gobble}%
\providecommand \bibinfo  [0]{\@secondoftwo}%
\providecommand \bibfield  [0]{\@secondoftwo}%
\providecommand \translation [1]{[#1]}%
\providecommand \BibitemOpen [0]{}%
\providecommand \bibitemStop [0]{}%
\providecommand \bibitemNoStop [0]{.\EOS\space}%
\providecommand \EOS [0]{\spacefactor3000\relax}%
\providecommand \BibitemShut  [1]{\csname bibitem#1\endcsname}%
\let\auto@bib@innerbib\@empty
\bibitem [{\citenamefont {Cao}\ \emph {et~al.}(2018)\citenamefont {Cao},
  \citenamefont {Romero},\ and\ \citenamefont
  {Aspuru-Guzik}}]{cao2018potential}%
  \BibitemOpen
  \bibfield  {author} {\bibinfo {author} {\bibfnamefont {Y.}~\bibnamefont
  {Cao}}, \bibinfo {author} {\bibfnamefont {J.}~\bibnamefont {Romero}},\ and\
  \bibinfo {author} {\bibfnamefont {A.}~\bibnamefont {Aspuru-Guzik}},\
  }\bibfield  {title} {\bibinfo {title} {Potential of quantum computing for
  drug discovery},\ }\href@noop {} {\bibfield  {journal} {\bibinfo  {journal}
  {IBM Journal of Research and Development}\ }\textbf {\bibinfo {volume}
  {62}},\ \bibinfo {pages} {6} (\bibinfo {year} {2018})}\BibitemShut {NoStop}%
\bibitem [{\citenamefont {Sadybekov}\ and\ \citenamefont
  {Katritch}(2023)}]{sadybekov2023computational}%
  \BibitemOpen
  \bibfield  {author} {\bibinfo {author} {\bibfnamefont {A.~V.}\ \bibnamefont
  {Sadybekov}}\ and\ \bibinfo {author} {\bibfnamefont {V.}~\bibnamefont
  {Katritch}},\ }\bibfield  {title} {\bibinfo {title} {Computational approaches
  streamlining drug discovery},\ }\href@noop {} {\bibfield  {journal} {\bibinfo
   {journal} {Nature}\ }\textbf {\bibinfo {volume} {616}},\ \bibinfo {pages}
  {673} (\bibinfo {year} {2023})}\BibitemShut {NoStop}%
\bibitem [{\citenamefont {Jayatunga}\ \emph {et~al.}(2022)\citenamefont
  {Jayatunga}, \citenamefont {Xie}, \citenamefont {Ruder}, \citenamefont
  {Schulze},\ and\ \citenamefont {Meier}}]{jayatunga2022ai}%
  \BibitemOpen
  \bibfield  {author} {\bibinfo {author} {\bibfnamefont {M.~K.}\ \bibnamefont
  {Jayatunga}}, \bibinfo {author} {\bibfnamefont {W.}~\bibnamefont {Xie}},
  \bibinfo {author} {\bibfnamefont {L.}~\bibnamefont {Ruder}}, \bibinfo
  {author} {\bibfnamefont {U.}~\bibnamefont {Schulze}},\ and\ \bibinfo {author}
  {\bibfnamefont {C.}~\bibnamefont {Meier}},\ }\bibfield  {title} {\bibinfo
  {title} {Ai in small-molecule drug discovery: A coming wave},\ }\href@noop {}
  {\bibfield  {journal} {\bibinfo  {journal} {Nat. Rev. Drug Discov}\ }\textbf
  {\bibinfo {volume} {21}},\ \bibinfo {pages} {175} (\bibinfo {year}
  {2022})}\BibitemShut {NoStop}%
\bibitem [{\citenamefont {Sabe}\ \emph {et~al.}(2021)\citenamefont {Sabe},
  \citenamefont {Ntombela}, \citenamefont {Jhamba}, \citenamefont {Maguire},
  \citenamefont {Govender}, \citenamefont {Naicker},\ and\ \citenamefont
  {Kruger}}]{sabe2021current}%
  \BibitemOpen
  \bibfield  {author} {\bibinfo {author} {\bibfnamefont {V.~T.}\ \bibnamefont
  {Sabe}}, \bibinfo {author} {\bibfnamefont {T.}~\bibnamefont {Ntombela}},
  \bibinfo {author} {\bibfnamefont {L.~A.}\ \bibnamefont {Jhamba}}, \bibinfo
  {author} {\bibfnamefont {G.~E.}\ \bibnamefont {Maguire}}, \bibinfo {author}
  {\bibfnamefont {T.}~\bibnamefont {Govender}}, \bibinfo {author}
  {\bibfnamefont {T.}~\bibnamefont {Naicker}},\ and\ \bibinfo {author}
  {\bibfnamefont {H.~G.}\ \bibnamefont {Kruger}},\ }\bibfield  {title}
  {\bibinfo {title} {Current trends in computer aided drug design and a
  highlight of drugs discovered via computational techniques: A review},\
  }\href@noop {} {\bibfield  {journal} {\bibinfo  {journal} {European Journal
  of Medicinal Chemistry}\ }\textbf {\bibinfo {volume} {224}},\ \bibinfo
  {pages} {113705} (\bibinfo {year} {2021})}\BibitemShut {NoStop}%
\bibitem [{\citenamefont {Morris}\ and\ \citenamefont
  {Lim-Wilby}(2008)}]{morris2008molecular}%
  \BibitemOpen
  \bibfield  {author} {\bibinfo {author} {\bibfnamefont {G.~M.}\ \bibnamefont
  {Morris}}\ and\ \bibinfo {author} {\bibfnamefont {M.}~\bibnamefont
  {Lim-Wilby}},\ }\bibfield  {title} {\bibinfo {title} {Molecular docking},\
  }\href@noop {} {\bibfield  {journal} {\bibinfo  {journal} {Molecular modeling
  of proteins}\ ,\ \bibinfo {pages} {365}} (\bibinfo {year}
  {2008})}\BibitemShut {NoStop}%
\bibitem [{\citenamefont {Bishwas}\ \emph {et~al.}(2017)\citenamefont
  {Bishwas}, \citenamefont {Mani},\ and\ \citenamefont
  {Palade}}]{bishwas2017quantum}%
  \BibitemOpen
  \bibfield  {author} {\bibinfo {author} {\bibfnamefont {A.~K.}\ \bibnamefont
  {Bishwas}}, \bibinfo {author} {\bibfnamefont {A.}~\bibnamefont {Mani}},\ and\
  \bibinfo {author} {\bibfnamefont {V.}~\bibnamefont {Palade}},\ }\bibfield
  {title} {\bibinfo {title} {Quantum algorithm for sequence clustering},\ }in\
  \href@noop {} {\emph {\bibinfo {booktitle} {Hybrid Intelligent Techniques for
  Pattern Analysis and Understanding}}},\ \bibinfo {editor} {edited by\
  \bibinfo {editor} {\bibfnamefont {M.~A. P. I. D.~P.}\ \bibnamefont
  {Bhattacharyya}, \bibfnamefont {S.}}\ and\ \bibinfo {editor} {\bibfnamefont
  {A.}~\bibnamefont {Bhaumik}}}\ (\bibinfo  {publisher} {Chapman and
  Hall/CRC},\ \bibinfo {address} {New York},\ \bibinfo {year} {2017})\ pp.\
  \bibinfo {pages} {345--364}\BibitemShut {NoStop}%
\bibitem [{\citenamefont {Pitchai}(2022)}]{pitchai2022quantum}%
  \BibitemOpen
  \bibfield  {author} {\bibinfo {author} {\bibfnamefont {A.}~\bibnamefont
  {Pitchai}},\ }\bibfield  {title} {\bibinfo {title} {Quantum computing:
  Principles and mathematical models},\ }\href@noop {} {\bibfield  {journal}
  {\bibinfo  {journal} {Emerging Computing Paradigms: Principles, Advances and
  Applications}\ ,\ \bibinfo {pages} {41}} (\bibinfo {year}
  {2022})}\BibitemShut {NoStop}%
\bibitem [{\citenamefont {Bishwas}\ \emph
  {et~al.}(2020{\natexlab{a}})\citenamefont {Bishwas}, \citenamefont {Mani},\
  and\ \citenamefont {Palade}}]{bishwas2020investigation}%
  \BibitemOpen
  \bibfield  {author} {\bibinfo {author} {\bibfnamefont {A.~K.}\ \bibnamefont
  {Bishwas}}, \bibinfo {author} {\bibfnamefont {A.}~\bibnamefont {Mani}},\ and\
  \bibinfo {author} {\bibfnamefont {V.}~\bibnamefont {Palade}},\ }\bibfield
  {title} {\bibinfo {title} {An investigation on support vector clustering for
  big data in quantum paradigm},\ }\href@noop {} {\bibfield  {journal}
  {\bibinfo  {journal} {Quantum Information Processing}\ }\textbf {\bibinfo
  {volume} {19}},\ \bibinfo {pages} {108} (\bibinfo {year}
  {2020}{\natexlab{a}})}\BibitemShut {NoStop}%
\bibitem [{\citenamefont {Bishwas}\ \emph
  {et~al.}(2020{\natexlab{b}})\citenamefont {Bishwas}, \citenamefont {Mani},\
  and\ \citenamefont {Palade}}]{bishwas2020gaussian}%
  \BibitemOpen
  \bibfield  {author} {\bibinfo {author} {\bibfnamefont {A.~K.}\ \bibnamefont
  {Bishwas}}, \bibinfo {author} {\bibfnamefont {A.}~\bibnamefont {Mani}},\ and\
  \bibinfo {author} {\bibfnamefont {V.}~\bibnamefont {Palade}},\ }\bibfield
  {title} {\bibinfo {title} {Gaussian kernel in quantum learning},\ }\href@noop
  {} {\bibfield  {journal} {\bibinfo  {journal} {International Journal of
  Quantum Information}\ }\textbf {\bibinfo {volume} {18}},\ \bibinfo {pages}
  {2050006} (\bibinfo {year} {2020}{\natexlab{b}})}\BibitemShut {NoStop}%
\bibitem [{\citenamefont {Heifetz}(2020)}]{heifetz2020quantum}%
  \BibitemOpen
  \bibfield  {author} {\bibinfo {author} {\bibfnamefont {A.}~\bibnamefont
  {Heifetz}},\ }\href@noop {} {\emph {\bibinfo {title} {Quantum mechanics in
  drug discovery}}}\ (\bibinfo  {publisher} {Springer},\ \bibinfo {address}
  {New York},\ \bibinfo {year} {2020})\BibitemShut {NoStop}%
\bibitem [{\citenamefont {Bishwas}\ \emph
  {et~al.}(2020{\natexlab{c}})\citenamefont {Bishwas}, \citenamefont {Mani},\
  and\ \citenamefont {Palade}}]{bishwas20204}%
  \BibitemOpen
  \bibfield  {author} {\bibinfo {author} {\bibfnamefont {A.~K.}\ \bibnamefont
  {Bishwas}}, \bibinfo {author} {\bibfnamefont {A.}~\bibnamefont {Mani}},\ and\
  \bibinfo {author} {\bibfnamefont {V.}~\bibnamefont {Palade}},\ }\bibfield
  {title} {\bibinfo {title} {4 from classical to quantum machine learning},\
  }\href@noop {} {\bibfield  {journal} {\bibinfo  {journal} {Quantum Machine
  Learning}\ }\textbf {\bibinfo {volume} {6}},\ \bibinfo {pages} {67} (\bibinfo
  {year} {2020}{\natexlab{c}})}\BibitemShut {NoStop}%
\bibitem [{\citenamefont {Hauke}\ \emph {et~al.}(2020)\citenamefont {Hauke},
  \citenamefont {Katzgraber}, \citenamefont {Lechner}, \citenamefont
  {Nishimori},\ and\ \citenamefont {Oliver}}]{hauke2020perspectives}%
  \BibitemOpen
  \bibfield  {author} {\bibinfo {author} {\bibfnamefont {P.}~\bibnamefont
  {Hauke}}, \bibinfo {author} {\bibfnamefont {H.~G.}\ \bibnamefont
  {Katzgraber}}, \bibinfo {author} {\bibfnamefont {W.}~\bibnamefont {Lechner}},
  \bibinfo {author} {\bibfnamefont {H.}~\bibnamefont {Nishimori}},\ and\
  \bibinfo {author} {\bibfnamefont {W.~D.}\ \bibnamefont {Oliver}},\ }\bibfield
   {title} {\bibinfo {title} {Perspectives of quantum annealing: Methods and
  implementations},\ }\href@noop {} {\bibfield  {journal} {\bibinfo  {journal}
  {Reports on Progress in Physics}\ }\textbf {\bibinfo {volume} {83}},\
  \bibinfo {pages} {054401} (\bibinfo {year} {2020})}\BibitemShut {NoStop}%
\bibitem [{\citenamefont {Zinner}\ \emph {et~al.}(2021)\citenamefont {Zinner},
  \citenamefont {Dahlhausen}, \citenamefont {Boehme}, \citenamefont {Ehlers},
  \citenamefont {Bieske},\ and\ \citenamefont {Fehring}}]{zinner2021quantum}%
  \BibitemOpen
  \bibfield  {author} {\bibinfo {author} {\bibfnamefont {M.}~\bibnamefont
  {Zinner}}, \bibinfo {author} {\bibfnamefont {F.}~\bibnamefont {Dahlhausen}},
  \bibinfo {author} {\bibfnamefont {P.}~\bibnamefont {Boehme}}, \bibinfo
  {author} {\bibfnamefont {J.}~\bibnamefont {Ehlers}}, \bibinfo {author}
  {\bibfnamefont {L.}~\bibnamefont {Bieske}},\ and\ \bibinfo {author}
  {\bibfnamefont {L.}~\bibnamefont {Fehring}},\ }\bibfield  {title} {\bibinfo
  {title} {Quantum computing's potential for drug discovery: Early stage
  industry dynamics},\ }\href@noop {} {\bibfield  {journal} {\bibinfo
  {journal} {Drug Discovery Today}\ }\textbf {\bibinfo {volume} {26}},\
  \bibinfo {pages} {1680} (\bibinfo {year} {2021})}\BibitemShut {NoStop}%
\bibitem [{\citenamefont {Gadioli}\ \emph {et~al.}(2021)\citenamefont
  {Gadioli}, \citenamefont {Palermo}, \citenamefont {Cherubin}, \citenamefont
  {Vitali}, \citenamefont {Agosta}, \citenamefont {Manelfi}, \citenamefont
  {Beccari}, \citenamefont {Cavazzoni}, \citenamefont {Sanna},\ and\
  \citenamefont {Silvano}}]{gadioli2021tunable}%
  \BibitemOpen
  \bibfield  {author} {\bibinfo {author} {\bibfnamefont {D.}~\bibnamefont
  {Gadioli}}, \bibinfo {author} {\bibfnamefont {G.}~\bibnamefont {Palermo}},
  \bibinfo {author} {\bibfnamefont {S.}~\bibnamefont {Cherubin}}, \bibinfo
  {author} {\bibfnamefont {E.}~\bibnamefont {Vitali}}, \bibinfo {author}
  {\bibfnamefont {G.}~\bibnamefont {Agosta}}, \bibinfo {author} {\bibfnamefont
  {C.}~\bibnamefont {Manelfi}}, \bibinfo {author} {\bibfnamefont {A.~R.}\
  \bibnamefont {Beccari}}, \bibinfo {author} {\bibfnamefont {C.}~\bibnamefont
  {Cavazzoni}}, \bibinfo {author} {\bibfnamefont {N.}~\bibnamefont {Sanna}},\
  and\ \bibinfo {author} {\bibfnamefont {C.}~\bibnamefont {Silvano}},\
  }\bibfield  {title} {\bibinfo {title} {Tunable approximations to control
  time-to-solution in an hpc molecular docking mini-app},\ }\href@noop {}
  {\bibfield  {journal} {\bibinfo  {journal} {The Journal of Supercomputing}\
  }\textbf {\bibinfo {volume} {77}},\ \bibinfo {pages} {841} (\bibinfo {year}
  {2021})}\BibitemShut {NoStop}%
\bibitem [{\citenamefont {Kochenberger}\ \emph {et~al.}(2014)\citenamefont
  {Kochenberger}, \citenamefont {Hao}, \citenamefont {Glover}, \citenamefont
  {Lewis}, \citenamefont {L{\"u}}, \citenamefont {Wang},\ and\ \citenamefont
  {Wang}}]{kochenberger2014unconstrained}%
  \BibitemOpen
  \bibfield  {author} {\bibinfo {author} {\bibfnamefont {G.}~\bibnamefont
  {Kochenberger}}, \bibinfo {author} {\bibfnamefont {J.-K.}\ \bibnamefont
  {Hao}}, \bibinfo {author} {\bibfnamefont {F.}~\bibnamefont {Glover}},
  \bibinfo {author} {\bibfnamefont {M.}~\bibnamefont {Lewis}}, \bibinfo
  {author} {\bibfnamefont {Z.}~\bibnamefont {L{\"u}}}, \bibinfo {author}
  {\bibfnamefont {H.}~\bibnamefont {Wang}},\ and\ \bibinfo {author}
  {\bibfnamefont {Y.}~\bibnamefont {Wang}},\ }\bibfield  {title} {\bibinfo
  {title} {The unconstrained binary quadratic programming problem: a survey},\
  }\href@noop {} {\bibfield  {journal} {\bibinfo  {journal} {Journal of
  combinatorial optimization}\ }\textbf {\bibinfo {volume} {28}},\ \bibinfo
  {pages} {58} (\bibinfo {year} {2014})}\BibitemShut {NoStop}%
\bibitem [{\citenamefont {Domino}\ \emph {et~al.}(2022)\citenamefont {Domino},
  \citenamefont {Kundu}, \citenamefont {Salehi},\ and\ \citenamefont
  {Krawiec}}]{domino2022quadratic}%
  \BibitemOpen
  \bibfield  {author} {\bibinfo {author} {\bibfnamefont {K.}~\bibnamefont
  {Domino}}, \bibinfo {author} {\bibfnamefont {A.}~\bibnamefont {Kundu}},
  \bibinfo {author} {\bibfnamefont {{\"O}.}~\bibnamefont {Salehi}},\ and\
  \bibinfo {author} {\bibfnamefont {K.}~\bibnamefont {Krawiec}},\ }\bibfield
  {title} {\bibinfo {title} {Quadratic and higher-order unconstrained binary
  optimization of railway rescheduling for quantum computing},\ }\href@noop {}
  {\bibfield  {journal} {\bibinfo  {journal} {Quantum Information Processing}\
  }\textbf {\bibinfo {volume} {21}},\ \bibinfo {pages} {337} (\bibinfo {year}
  {2022})}\BibitemShut {NoStop}%
\bibitem [{\citenamefont {Mandal}\ \emph {et~al.}(2020)\citenamefont {Mandal},
  \citenamefont {Roy}, \citenamefont {Upadhyay},\ and\ \citenamefont
  {Ushijima-Mwesigwa}}]{mandal2020compressed}%
  \BibitemOpen
  \bibfield  {author} {\bibinfo {author} {\bibfnamefont {A.}~\bibnamefont
  {Mandal}}, \bibinfo {author} {\bibfnamefont {A.}~\bibnamefont {Roy}},
  \bibinfo {author} {\bibfnamefont {S.}~\bibnamefont {Upadhyay}},\ and\
  \bibinfo {author} {\bibfnamefont {H.}~\bibnamefont {Ushijima-Mwesigwa}},\
  }\bibfield  {title} {\bibinfo {title} {Compressed quadratization of higher
  order binary optimization problems},\ }in\ \href@noop {} {\emph {\bibinfo
  {booktitle} {Proceedings of the 17th ACM International Conference on
  Computing Frontiers}}}\ (\bibinfo {year} {2020})\ pp.\ \bibinfo {pages}
  {126--131}\BibitemShut {NoStop}%
\bibitem [{wwp(2019)}]{wwpdb2019protein}%
  \BibitemOpen
  \bibfield  {title} {\bibinfo {title} {Protein data bank: the single global
  archive for 3d macromolecular structure data},\ }\href@noop {} {\bibfield
  {journal} {\bibinfo  {journal} {Nucleic acids research}\ }\textbf {\bibinfo
  {volume} {47}},\ \bibinfo {pages} {D520} (\bibinfo {year}
  {2019})}\BibitemShut {NoStop}%
\bibitem [{\citenamefont {Mato}\ \emph {et~al.}(2022)\citenamefont {Mato},
  \citenamefont {Mengoni}, \citenamefont {Ottaviani},\ and\ \citenamefont
  {Palermo}}]{mato2022quantum}%
  \BibitemOpen
  \bibfield  {author} {\bibinfo {author} {\bibfnamefont {K.}~\bibnamefont
  {Mato}}, \bibinfo {author} {\bibfnamefont {R.}~\bibnamefont {Mengoni}},
  \bibinfo {author} {\bibfnamefont {D.}~\bibnamefont {Ottaviani}},\ and\
  \bibinfo {author} {\bibfnamefont {G.}~\bibnamefont {Palermo}},\ }\bibfield
  {title} {\bibinfo {title} {Quantum molecular unfolding},\ }\href@noop {}
  {\bibfield  {journal} {\bibinfo  {journal} {Quantum Science and Technology}\
  }\textbf {\bibinfo {volume} {7}},\ \bibinfo {pages} {035020} (\bibinfo {year}
  {2022})}\BibitemShut {NoStop}%
\bibitem [{\citenamefont {Sax}\ \emph {et~al.}(2020)\citenamefont {Sax},
  \citenamefont {Feld}, \citenamefont {Zielinski}, \citenamefont {Gabor},
  \citenamefont {Linnhoff-Popien},\ and\ \citenamefont
  {Mauerer}}]{sax2020approximate}%
  \BibitemOpen
  \bibfield  {author} {\bibinfo {author} {\bibfnamefont {I.}~\bibnamefont
  {Sax}}, \bibinfo {author} {\bibfnamefont {S.}~\bibnamefont {Feld}}, \bibinfo
  {author} {\bibfnamefont {S.}~\bibnamefont {Zielinski}}, \bibinfo {author}
  {\bibfnamefont {T.}~\bibnamefont {Gabor}}, \bibinfo {author} {\bibfnamefont
  {C.}~\bibnamefont {Linnhoff-Popien}},\ and\ \bibinfo {author} {\bibfnamefont
  {W.}~\bibnamefont {Mauerer}},\ }\bibfield  {title} {\bibinfo {title}
  {Approximate approximation on a quantum annealer},\ }in\ \href@noop {} {\emph
  {\bibinfo {booktitle} {Proceedings of the 17th ACM International Conference
  on Computing Frontiers}}}\ (\bibinfo {year} {2020})\ pp.\ \bibinfo {pages}
  {108--117}\BibitemShut {NoStop}%
\bibitem [{\citenamefont {Van~Verth}\ and\ \citenamefont
  {Bishop}(2015)}]{van2015essential}%
  \BibitemOpen
  \bibfield  {author} {\bibinfo {author} {\bibfnamefont {J.~M.}\ \bibnamefont
  {Van~Verth}}\ and\ \bibinfo {author} {\bibfnamefont {L.~M.}\ \bibnamefont
  {Bishop}},\ }\href@noop {} {\emph {\bibinfo {title} {Essential mathematics
  for games and interactive applications}}}\ (\bibinfo  {publisher} {CRC
  Press},\ \bibinfo {address} {Boca Raton},\ \bibinfo {year}
  {2015})\BibitemShut {NoStop}%
\end{thebibliography}%

\end{document}